# Optimal Routing and Scheduling of Charge for Electric Vehicles: Case Study[1]

J. Barco[1], A. Guerra[2], L. Muñoz[2], N. Quijano[1]

*Universidad de los Andes.* [1]*Electrical and Electronics Engineering Department.* [2]*Mechanical Engineering Department. Bogotá, Colombia*

**Abstract**

In Colombia, there is an increasing interest about improving public transportation. One of the proposed strategies in that way is the use battery electric vehicles (BEVs). One of the new challenges is the BEVs routing problem, which is subjected to the traditional issues of the routing problems, and must also consider the particularities of autonomy, charge and battery degradation of the BEVs. In this work, a scheme that coordinates the routing, scheduling of charge and operating costs of BEVs is proposed. The simplified operating costs have been modeled considering both charging fees and battery degradation. A case study is presented, in order to illustrate the proposed methodology. The given case considers an airport shuttle service scenario, in which energy consumption of the BEVs is estimated based on experimentally measured driving patterns.

## 1. Introduction

The integration of battery electric vehicles (BEVs) into the public transport sector has been encouraged [1] due to two main reasons: Energy efficiency and reduction of the carbon footprint. From the energetic point of view, the energy conversion since EVs use the energy more efficiently, the charging of an EV is cheaper than the fuelling internal combustion engine vehicles (ICEVs). From the emissions point of view, when BEVs are used in combination with clean sources for the generation of electricity the result is a reduction of the emissions associated to the combustion of fossil fuels. Therefore, the BEVs are one of the best alternatives to be integrated into the cities as a public means of transport.

The implementation of BEVs in the public transport service involves several challenges associated to the combination of the characteristics of the service with those of the vehicles. Specially, three issues are highlighted: routing of EVs' dedicated to public transport, the scheduling of charge, and the battery health.

First of all, in order to be economically and technically feasible, the BEVs must be routed to minimize the energy consumption. To achieve this, the following two steps are considered: i) it is necessary to find the minimum consumption paths to travel between two points. For this minimization, the particular characteristic of BEVs must be taken into account, (e.g., the BEVs ability to recover energy by traveling in a way that can take advantage of the regenerative brake; ii) the optimal routes (composed by minimum consumption paths) must be determined to meet the transport demand in different places at different times while the energy consumption is minimized. The computation of the routing scheme of BEVs must be made taking into account the range of the given vehicles, imposed mainly by the available battery technologies [2]. This consideration may require intermediate charging stages of the vehicles to extend their range.

[*] This work has been supported partially by project CIFI 2012, Facultad de Ingeniería Universidad de los Andes, and also by Codensa S.A. ESP trough Silice III project, under Grant P12.245422.006.2012.
Email addresses: je.barco69@uniandes.edu.co, af.guerra69@uniandes.edu.co, lui-muno@uniandes.edu. co, nquijano@uniandes.edu.co

Second, scheduling of charge must be coordinated with routing to guarantee a reliable operation while the cost of charging is minimized. In this way, the scheduling of charge must be made considering the variation of the energy tariff during the peak and non-peak hours of the day. Furthermore, the scheduling of recharge must take into account which is the quantity of energy required to perform the next travel, and the recharging time required to perform the recharge.

Finally, the battery health is considered, as it is sensitive to charging/discharging actions [3]. Also, the battery is the most expensive component of the BEV [4]. Hence, the planning of a charge schedule could help to increase the battery lifetime reducing operational costs.

In the literature, there are several papers discussing the three topics of interest introduced above. However, most of the studies deal only with one issue at a time. In [5, 6] the energy consumption by travel is discussed. A reduced order model of energy consumption is proposed, which considers characteristics of the BEVs (weight, rolling resistance, and drag coefficients) and the characteristics of the road (grade, distance, and traffic speed). From this model, an energy graph can be constructed which allows to find the minimum energy path between two points using classical algorithms (e.g., Dijkstra [7], Johnson [8]).

In [6, 9] the *electric vehicle routing problem* (EVRP) formulated consists on finding a set of minimum cost routes, such that the demand of the costumer is satisfied. In addition, some new constraints are introduced to take into account the capacity of the battery of the BEVs. In [10 - 12] the model of EVRP also considers the refuelling and the recharging time of BEVs. Nevertheless, the costs of recharge and battery degradation have not been taken into account in these works.

Other papers such as [13 - 16] discuss the recharging control in order to minimize the cost of energy recharge. Both works [14, 16] address the recharging control for large fleets of BEVs. The control strategy obtained minimizes generation costs and the recharge cost of the BEVs fleet. Also, it highlights that the strategy is not centralized, allowing to preserve the autonomy of each vehicle. However, these works do not consider the energy consumed by each BEV and the fact that the optimal charging profiles obtained may lead to an increased degradation of the battery.

In [17 - 19] another approach to obtain BEVs charging profiles is discussed. An optimization problem is formulated to minimize power losses and maximize the load factor. The formulation takes into account a detailed topology and network characteristics. In this way, the charging profiles obtained indicate that, BEVs plugged in points that could induce large losses must be charged in valley hours. Nevertheless, these studies do not consider the effect of charging profile over battery life nor the cost of recharging in minimizing the objective function.

In [20, 21] battery *state of health* (SOH) is studied using a battery degradation model for lithium-ion batteries. The model considers the main features that influence the degradation of the battery such as: *state of charge* (SOC) average, *depth of discharge* (DOD), and battery temperature. In [20] the models used to minimize the cost of battery degradation and the cost of energy recharge. The charging profile obtained shows a considerable difference from the one obtained in [14, 16], where the recharge profile is calculated taking into account only the cost of recharging.

This work proposes a methodology that relates the three topics listed above to determine: the minimum consumption paths, the set of optimal routes, the routes assignment between the vehicles, and the scheduling of recharge for public transport EVs. A centralized controller based on a program that minimizes the operation cost of all EVs is proposed. The objective function to be minimized considers both the recharging cost and the battery degradation cost associated to the routes assignment and recharging actions.

As case study, the *vehicle routing problem* (VRP) associated to an airport shuttle service served by BEVs is considered. This service consists of carrying passengers from an airport to a hotel, while the aforementioned issues are considered. Additionally, we present some comparative results between the scheduling of recharge obtained using a reduced order model and real data of energy consumption of an electric vehicle. Finally, we present ideas on patterns of charging, routing, and operating costs of BEVs, which would be useful for public transport companies that are planning to upgrade their fleet with BEVs.

The remainder of this paper is organized as follows. Section 2 presents the approximated energy consumption model and the vehicle routing problem. In Section 3, we introduce the scheduling of recharge problem, which consists of routes assignation and charging model. Sections 4 and 5 describe the battery degradation model and an evolutive method to solve the proposed optimization problem. Finally, Sections 6 and 7 show some simulations results and concluding remarks.

## 2. Optimal Routing for EVs

The scenario under study considers that there is an *operation center* (OC) which coordinates both the transport service and the scheduling of charge. The transport service consists of a BEVs fleet that provides a shuttle airport-hotel (SA). On the other hand, the scheduling of charge of BEVs fleet is planned while the cost is minimized. Thus, the OC assigns routes to BEVs to meet the transport demand and also schedules the recharging actions in the available *recharging stations* (RSs) according to the tariff of energy. In order to coordinate the operation of BEVs fleet, the OC is able to communicate with BEVs, clients (potential passengers), RSs, and utility (this scenario is shown in Fig. 1). Thus, the OC receives information such as: requests of clients a day ahead, the availability in RSs, BEVs SOC, and the tariff of energy from utility. Then, the routes assignment and the scheduling of recharge can be calculated and broadcasted to the EVs. Next, we describe each model in which is based the working of the OC, which are presented as follows: estimation of energy consumed on the road, determination of routes, routes assignment and, finally we present the scheduling of recharge problem and its solution method namely differential evolution.

*2.1. Energy Consumption on the Road*

For BEVs, the energy consumption on the road is sensitive to environmental characteristics, road characteristics, as well as to the characteristics of the vehicle. Factors such as road elevation and traveling speed can significantly change the energy consumption. Traffic conditions and environmental factors induce acceleration/deceleration rates that have important impact on the vehicle energy consumption.

Therefore, a model that considers a vehicle moving at a constant speed is not sufficient.

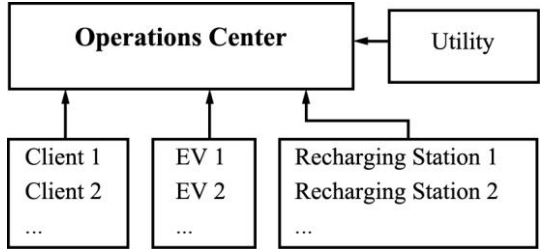

Figure 1. Scheme for BEV operation.

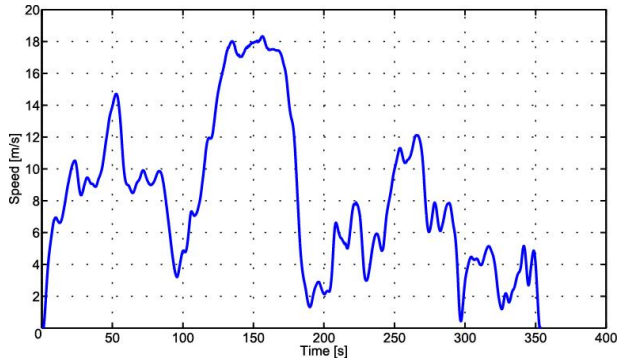

Figure 2. Example of a speed profile for one of the paths.

The energy consumption model presented in this section is based on the longitudinal dynamics equations of motion presented in [22], with a constant rolling resistance coefficient as a constant value as the vehicle speed changes. It considers only the energy consumed by the EV while it is accelerating or moving at a constant speed; it does not consider the energy recovered by the regenerative braking system of the vehicle. Also, it determines the energy consumption of the vehicle as a function of environmental characteristics, road characteristics and the characteristics of the vehicle.

Road characteristics are modeled in a directed graph $\mathcal{G} = (\mathcal{V}, \mathcal{E})$. Vertexes (key-points) $v \in \mathcal{V}$ represent points of special interest on the street map. Edges $\varepsilon_{ij} \in \mathcal{E}$, where $\varepsilon_{ij} = (v_i, v_j)$. The edges represent road sections between key-points. The model assumes that, for every edge, the representative driving pattern associated to the given edge is known and it is represented by its speed profile. Hence, there is a function $s: \mathcal{E} \to \kappa(\mathbb{R})$, where $s_{ij} = s(\varepsilon_{ij})$ corresponds to the representative speed profile associated to the road section that connects vertexes $v_i$ and $v_j$ and $\kappa(\mathbb{R})$ is the space of continuous real functions with compact support. The model also assumes that the key-point elevation $z: \mathcal{V} \to \mathbb{R}$ is given. Furthermore, the characteristics of the vehicle that influence the energy consumption are: mass (m), frontal area (A), drag and rolling resistance coefficients $(C_D, f_r)$, and power-train efficiency $(\hat{\eta})$.

Given these characteristics, the power consumption during the instant t of the travel between the vertex *i* and the vertex *j* is given by:

$$P_{i,j}(t) = \frac{P_{aerodynamic}(t) + P_{rolling}(t) + P_{specific}(t) + P_{gravitational}(t)}{\hat{\eta}} \quad (1)$$

$$P_{i,j}(t) = \frac{\frac{1}{2}\rho A C_D [s_{ij}(t)]^3 + mgf_r s_{ij}(t) + m\frac{ds_{ij}}{dt}(t)s_{ij}(t) + mg\tan(\gamma)s_{ij}(t)}{\hat{\eta}} \quad (2)$$

The first term on the right hand side of the equation corresponds for the power dissipated by the aerodynamic drag, the second for the rolling resistance between the tires and the asphalt, the third for the specific power required to overcome the inertia of the vehicle and the fourth corresponds to the power required due to the change the height of the vehicle originated by the grade of the road, represented by the grade angle g.

The power consumption can be obtained through the integration of Eq. (2) over the total time of each trip between vertexes in order to obtain the energy consumption associated to the trip. Under the assumption of constant power-train efficiency, the gravitational component can be analytically integrated, leading to the following expression:

$$c_{i,j} = \int_0^{t_f} \left( \frac{\frac{1}{2}\rho A C_D [s_{ij}(t)]^3 + mgf_r s_{ij}(t) + m\frac{ds_{ij}}{dt}(t)s_{ij}(t)}{\hat{\eta}} \right) dt - \frac{mg[z(v_j) - z(v_i)]}{\hat{\eta}} \quad (3)$$

Finally, the energy consumption can be generalized to an arbitrary path, composed by different edges of the graph. A path $P_h$ is defined, as a sequence of *l* vertexes $(v_1, v_2, \ldots, v_l)$ with $(v_i, v_{i+1}) \in \mathcal{E}$ for $i = 1, \ldots, l-1$. The energy needed to travel along a path $P_h$ in the road network is the sum of the energy consumed to complete each one of the road section that conform the path:

$$C(P_h) = \sum_{i=1}^{l-1} c_{i,j} \quad (4)$$

*2.2. Determination of Routes*

Similarly to the classical vehicle routing problems formulated in [23], we propose an EVRP to model the airport-hotel shuttle service. It consists on finding a set of minimum consumption routes, such that the transport demand and all operational constraints are satisfied. First, we show some notation used to explain the routing model in table 1.

The EVRP model is formulated on an energy graph $\mathcal{G}_s$, which is a simplification of the road network. Let $\mathcal{G}_s = (\mathcal{V}_s, \mathcal{E}_s)$ be a simplified energy graph. The nodes are composed by $\mathcal{V}_s = \{v_{d_1}, v_{d_2}\} \cup C \cup R_s$ where C denotes the set of nodes with transport demand, $R_s$ denotes the set of RSs. Also, we set $2C = |C|$ as the number of clients requests, where the operator $|\cdot|$ represents the modulo of a set. The depot nodes are denoted by $v_{d_1}$ and $v_{d_2}$, which represent the start and end nodes of a route. The edges are defined by $\mathcal{E}_s = \{(v_i, v_j) | v_i, v_j \in \mathcal{V}_S\}$ where $(v_i, v_j) \in \mathcal{E}_s$ are minimum consumption paths that can be found using routing algorithms on the road network model (this process is detailed above in section 2.1). On the other hand, the transport demand involves *m* requests with pickup and delivery stops as well as associated demands. The requests are identified by two nodes $i$ and $m + i$, corresponding to the pickup and delivery stops of request, respectively. The set of pickup nodes is denoted by $P = \{1, \dots, m\}$, and the set of delivery nodes is denoted by $D = \{m + 1, \dots, 2m\}$. In this way, we define $C = P \cup D$ and if request *i* consists of transporting qi passengers from $i$ to $m + i$, then $q_{m+i} = -q_i$.

| Variable | Description |
|---|---|
| $\mathcal{G}_s$ | Simplified energy graph $\mathcal{G}_s$. |
| $\mathcal{V}_s$ | Vertices or nodes of the energy graph $\mathcal{G}_s$. |
| $\mathcal{E}_s$ | Edges of the energy graph. |
| $v_{d_1}, v_{d_2}$ | Instances of depot node. |
| $C$ | Node set of clients. |
| $R_s$ | Node set of charging stations. |
| $P$ | Set of pick up nodes. |
| $D$ | Set of delivery nodes. |
| $c_{ij}$ | Energy consumed by travel from node *i* to *j*. |
| $t_{ij}$ | Time elapsed on a trip from node *i* to *j*. |
| $[a_i, b_i]$ | Time window during which passenger in node *i* must be picked up. |
| $q_i$ | Number of passengers in node *i*. |
| $Q$ | Maximum load capacity of passengers. |
| $B$ | Battery capacity. |
| $M$ | M is a constant with a value greater than any $w_j^h$. |
| $e_{min}$ | Lower bound of the battery level. |
| $H$ | Set of routes. |
| $x_{ij}^h$ | Binary flow variable which specifies the route *h* travel along node *i* to *j*. |
| $y_j^h$ | Amount of passengers picked up or delivered when traveling along a route *h* towards the node *j*. |
| $w_j^h$ | Time at which the passenger *i* is picked up or delivered. |
| $e_j^h$ | Remaining battery charge level at the end of route *h*, on vertex *i*. |

Table 1: Notation for the BEV routing problem.

The energy consumed by travel from *i* to *j* is given by $c_{ij}$, and its travel time is $t_{ij}$, where $i, j \in \mathcal{V}_s$. Each customer $i \in C$ has a time window $[a_i, b_i]$ during which the service must take place. Also, the EVs have a maximum load capacity *Q* and battery capacity *B*.

The objective function minimizes the energy consumption of all routes *H*, i.e.,

$$min \sum_{h \in H} \sum_{(i,j) \in \mathcal{E}_s} c_{ij} x_{ij}^h \qquad (5)$$

where h ∈ H is the set of routes to meet all transport demand, $x_{ij}^h$ are the flow variables, which are equal to 1 if arc (i, j) is used by route h and 0 otherwise. Also, we define two constraints to guarantee that all passenger demands are properly satisfied, which are

$$\sum_{h \in H} \sum_{j \in \mathcal{V}_s \setminus v_{d_1}} x_{ij}^h = 1 \quad \forall i \in P \tag{6}$$

$$\sum_{i \in \mathcal{V}_s \setminus v_{d_2}} x_{ij}^h - \sum_{i \in \mathcal{V}_s \setminus v_{d_1}} x_{j,m+1}^h = 0$$
$$\forall h \in H, j \in \mathcal{V}_s \setminus \{v_{d_1}, v_{d_2}\} \tag{7}$$

Eq. (6) guarantees that each node with passenger demands is attended by at most a route $h$. Moreover, Eq. (7) imposes that the passengers picked up in the nodes $i$ are delivered in the nodes $m+i$. We also define three constraints to satisfy the flow through vertices as

$$\sum_{i \in \mathcal{V}_s \setminus v_{d_2}} x_{ij}^h - \sum_{i \in \mathcal{V}_s \setminus v_{d_1}} x_{ji}^h = 0$$
$$\forall h \in H, j \in \mathcal{V}_s \setminus \{v_{d_1}, v_{d_2}\}, i \neq j \tag{8}$$

$$\sum_{j \in \mathcal{V}\_s \setminus v_{d_2}} x_{v_{d_1},j}^h \leq 1, \quad \forall h \in H \tag{9}$$

$$\sum_{i \in \mathcal{V}\_s \setminus v_{d_2}} x_{i,v_{d_2}}^h \leq 1, \quad \forall h \in H \tag{10}$$

Eq. (8) establishes that if a route $h$ enters to node $j$ then the same route $h$ must go out of node $j$. Besides, Eq. (9) indicates that at most one route $h$ could go out from the start depot vertex $v_{d_1}$, On the other hand, Eq. (10) indicates that at most one route $h$ could enter to the end depot vertex $v_{d_2}$. Furthermore, three sets of constraints are defined: i) the time constraints; ii) the capacity constraints; and iii) the energy constraints. Capacity constraints are

$$y_j^h \geq y_i^h + q_j - Q + Q x_{ij}^h, \quad \forall h \in H, i, j \in P, i \neq j \tag{11}$$

$$q_j \leq y_j^h \leq Q, \quad \forall h \in H, j \in \mathcal{V}_s \tag{12}$$

$$0 \leq y_{j+m}^h \leq Q - q_j, \quad \forall h \in H, j \in P \tag{13}$$

where $y_j^h$ is the capacity variable representing the amount of passengers picked or delivered by traveling along the route $h$ up to the node $j$. Therefore, Eq. (11) counts the number of passengers traveling along the route $h$ and constraints (12) and (13) guarantee that the capacity variable does not exceed the maximum capacity $Q$. Time constraints are

$$w_j^h \geq w_i^h + t_{ij} - M + M x_{ij}^h, \quad \forall h \in H, i, j \in \mathcal{V}_s, i \neq j \tag{14}$$

$$a_j \leq w_j^h \leq b_j, \quad \forall h \in H, j \in \mathcal{V}_s \tag{15}$$

Where $w_i^h$ is the time variable that specifies the instant at the passenger $i$ is pick up or delivery and $M$ is a constant with a value greater than any $w_i^h$. Eq. (14) counts the time elapsed until that the passenger *i* is attended by the route *h*. In addition, the constraint in Eq. (15) guarantees that the passenger *i* is picked up or delivered within his time window. Finally, energy constraints are

$$e_j^h \leq e_i^h - c_{ij} x_{ij}^h + B - B x_{ij}^h, \qquad \forall h \in H, i,j \in \mathcal{V}_s, i \neq j \tag{16}$$

$$e_{\min} \leq e_j^h, \qquad \forall h \in H, j \in \mathcal{V}_s \tag{17}$$

Where $e_i^h$ is the energy variable that specifies the remaining battery charge level when the route *h* arrives to vertex *i*. Eq. (16) tracks the battery level based on vertices sequence, which is reduced based on the energy consumed for travel from vertex *i* to *j*. Also, Eq. (17) guarantees that the level of the battery is never below the minimum level $e_{\min}$.

The solution of the proposed EVPR model is a set of optimal routes *H*, which minimizes the energy consumed by satisfying a passenger demand, in this way the cost of charge is also indirectly reduced.

## 3. Scheduling of Charge Problem

The problem of scheduling of charge for BEVs is now introduced. First, some notation used in this section is presented in Table 2. Second, the assignment of routes between the fleet of BEVs is defined. Third, the model of charging control is presented, and finally the problem formulation is shown.

### 3.1. Assignment of Routes

The assignment of routes is the distribution of all optimal routes *S* between the BEVs fleet. The assignment of a set of routes between a fleet of *K* vehicles over a programming horizon *N* is considered. The programming horizon is shown in Fig. 3. The assignation is denoted by $a_s^k$, which is a binary variable that is equal to 1 if route *s* is assigned to EV, $k \in K$, and 0 otherwise. Furthermore, a binary variable named *unavailability* is defined. This variable indicates which periods of *i* E *N* are used in travelling along the route *s* assigned to an EV *k*. The unavailability variable and it is defined as

$$d_s^k(i) = \begin{cases} 1, & if \ t_s^s \leq i \leq t_s^e \\ 0, & otherwise \end{cases} \tag{18}$$

$$d_{s1}^k(i) + d_{s2}^k(i) \leq 1, \qquad \forall i \in N, \qquad \forall k \in K, s_1, s_2 \in S \tag{19}$$

| Variable | Description |
|---|---|
| $\mathcal{N}$ | Programming horizon. |
| $N$ | Final time slot of programming horizon. |
| $a_k^h$ | Route assignation variable. |
| $d_h^k(n)$ | Unavailability variable. |
| $t_h^s$ | Start time for a trip on route $h$. |
| $t_h^e$ | End time for a trip on route $h$. |
| $\boldsymbol{a}$ | Assignation profile. |
| $SOC^k(n)$ | State of charge of the BEV batteries $k$ at instant $n$. |
| $\eta_x$ | Charge efficiency of the charging station $x$. |
| $r_x$ | Charging rate of the station $x$. |
| $u_x^k(n)$ | Charging action variable of BEV batteries $k$ at instant $n$. |
| $e_h^k(n)$ | Consumed energy variable of BEV $k$ at instant $n$. |
| $\boldsymbol{u}^k$ | Charging profile for the batteries $k$. |
| $p_x(n)$ | Energy price in the recharging station $x$ at instant $n$. |
| $c_{deg}^k$ | Battery degradation cost $k$. |

Table 2: Notation for the scheduling of charge problem.

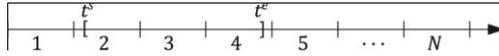

Figure 3: Timeline of programming horizon.

where $[t_s^s, t_e^s]$ is the interval duration of each route $s$, with start time $t_s^s$ and end time $t_e^s$. In this way, the variable $d_s^k(i)$ allows to locate each route $s$ on the horizon $N$ and Eq. (19) guarantees that two routes $(s_i, s_2)$ with similar duration intervals are not assigned to same BEV. Also, the energy consumed by BEV $k$ in the route $s$ on the interval $i$ of the horizon $N$ is defined as $e_k^s(i)$.

Finally, we define a profile of assignment by the vector $\boldsymbol{a} = \{a_s^k : \sum_{k \in K} \sum_{s \in S} a_s^k = |S|$ where the double summation guarantees that all routes $S$ are assigned between all EVs. Thus, the profile $\boldsymbol{a}$ is valid if the constraint in Eq. (19) is satisfied for the BEVs fleet.

### 3.2. Model of Charging Control

Taking some ideas from [14], we consider the charging control for an EVs fleet $K$ over a programming horizon $N$. It defines the state of charge of a vehicle $k \in K$ at instant $i \in N$ as $SOC^k(i)$, and the SOC dynamics are

$$SOC^k(i) = SOC^k(i-1) + \eta_x r_x u_x^k(i) - e_s^k(i) - e_x^k \\ \forall i \in N, \quad \forall k \in K \tag{20}$$

where $u_x^k(i)$ is a binary variable, which is equal to 1 when a charging action of vehicle $k$ at instant $i$ is performed, $r_x$ is the charging rate, and $\eta_x$ correspond to the charge efficiency. The subscript $x \in R$ marks the spot where the charge is done. $e_x^k(i)$ is the energy consumed by travel to RS $x$, which is explained in the next subsection. Now, a constraint to bound the $SOC$ is defined by

$$SOC_{\min} \leq SOC^k(i) \leq SOC_{\max} \quad i \in N, \forall k \in K \tag{21}$$

where $SOC_{min}$ and $SOC_{max}$ are the minimum and maximum limits of SOC, respectively. Both limits minimum and maximum could be fixed by recommendation of the battery manufacturer. Next, we define a constraint to guarantee that the initial and final $SOCs$ are equal to $SOC_{max}$.

$$SOC^k(0) = SOC_{\max}, SOC^k(N) = SOC_{\max}, \quad \forall k \in K \tag{22}$$

where $SOC^k(0)$ and $SOC^k(N)$ are the initial and final $SOCs$ of an BEV $k$. In Eq. (22) the initial and final conditions of the $SOC$ are chosen equal to each other, to ensure that in the next programming horizon the initial condition is the same.

Finally, we define a charging profile of a vehicle $k$ as a set of charging actions $u_x^k(i)$. A charging profile of an EV $k$ is denoted by $\boldsymbol{u}^k$ and it is defined by $\boldsymbol{u}^k = \{u_x^k(i): \eta_x r_x \sum_{i \in N} u_x^k(i) = \sum_{i \in N}(e_s^k(i) + e_x^k(i))\}$, where the sum of all charging actions must be equal to the energy consumed in all travels. The profile $\boldsymbol{u}^k$ is valid if the constraints (22) to (22) are satisfied.

*3.3. Reroute to Charging Stations*

According to the model of determination of routes when a vehicle $k$ is not traveling along any route then, the vehicle is parked in the depot vertex $(v_{d_1}, v_{d_2})$. Also, it is assumed that there is an **RS** $R_1 \in R$ in the depot vertex where the vehicles can be charged. In this scenario, it is defined a reroute to an RS as a travel to another RS to perform an action of charging $u_x^k : x \neq R_1$. The reroutes are marked by the subscript x in the action of charging $u_x^k$, if subscript $x$ points a different place to $R_1$ then, the BEV $k$ must travel to another RS $R_2 \in R$ to perform its action of recharging.
The reroute to RS implies that the BEV must have available time to traveling back and forth to RS. Hence, The BEV must have a few intervals available before/after of recharging to travel from depot vertex to RS and return. Two constraints are defined to guarantee this. The first constraint is

$$d_s^k(j) + u_y^k(j) = 0$$
$$i_x - \Delta i_x \leq j < i_x, \quad \forall y \neq x, y \in R \tag{23}$$

where $i_x \in N$ is the start interval of the recharging $u_x^k$, $\Delta i_x$ is the number of time intervals that a BEV $k$ takes to travel from $v_{d_1}$ to x. The sum of terms equal to zero in Eq. (23) guarantees that the BEV $k$ is available to be rerouted to RS. The second constraint is

$$d_s^k(j) + u_y^k(j) = 0$$
$$i_x + \Delta i_x \leq j < i_x + \Delta i_{u_x} + \Delta i_{v_{d_1}}, \quad \forall y \neq x \tag{24}$$

where $\Delta i_{u_x}$ is the number of time intervals that the recharging action takes in $\Delta i_{v_{d_1}}$ is the number of time intervals that a BEV $k$ takes to travel from x to $v_{d_1}$. Both Eqs. (23) and (24) indicate that the BEV $k$ can only be rerouted when there is time availability. Furthermore, we denote the energy consumed in the reroute as $e_x^k$, which is the energy consumed to travel from $v_{d_1}$ to *x* and return.

*3.4. Optimization Problem of Charge for EVs*

The optimization problem of charge for the BEVs fleet is now formulated. The model proposed seeks to minimize the operation cost of EVs fleet, which is defined by

$$\min \sum_{k \in K} \sum_{i \in N} p_x(i) u_x^k(i) + \sum_{k \in K} c_{\text{deg}}^k \tag{25}$$

The first term is the cost of charging energy of all fleet of BEVs, where $p_x(i)$ is the price of energy in the recharging station $x$ in the instant $i$, and the last term is the cost of battery degradation of all fleet of BEVs. The objective (25) is subject to the assignment profile **a** and charging profile $\boldsymbol{u}^k$ for all $k \in K$. In addition, if there are reroutes to recharging stations then the constraints in Eqs. (23) and (24) must be satisfied.
Also, it is defined the availability in each RS by $\sum_{k \in K} u_x^k(i) \leq A_x(i)$, $\forall \in N, x \in R$, which guarantees that recharging actions are scheduled in an available recharge spots, where $A_x(i)$ is the available spots number in RS $x$.

## 4. Battery Degradation Model

A simplified version of the lithium-ion battery degradation model presented in [20] is presented. First, the notation used is presented in Table 3, then the degradation model is described.

| Variable | Description |
|---|---|
| $c_{bat}$ | Battery cost. |
| $L_{Q,T}$ | Lifetime degradation due to temperature. |
| $L_{Q,SOC}$ | Lifetime degradation due to state of charge. |
| $L_{Q.DOD}$ | Lifetime degradation due to depth of discharge. |
| $nhy$ | Number of hours in a year. |
| $l_y(\cdot)$ | Function of the lifespan of the battery [years]. |
| $T_{amb}$ | Ambient temperature. |
| $R_{th}$ | Thermal resistance. |
| $P_t(\cdot)$ | Power of charging. |
| $t_{max}$ | Available time to perform a charge [hours]. |
| $t_{ch}$ | Charge time [hours]. |
| $CF_{max}$ | Fading Capacity. |
| $y_p$ | Estimated lifespan for the batteries. |
| $N_l(\cdot)$ | Lifespan of battery in cycles. |

Table 3: Notation for the battery degradation model.

The model estimates the cost of battery degradation $C_{\text{deg}}$ in terms of battery life reduction. Thus, $C_{\text{deg}}$ is a function of three factors that could considerably reduce the battery life such as: temperature, SOC, and DOD. It is defined as follows

$$c_{\text{deg}} = c_{bat}(L_{Q,T} + L_{Q,SOC} + L_{Q,DOD}) \tag{26}$$

where $c_{bat}$ is the cost of battery, and $L_{Q,T}$, $L_{Q,SOC}$, $L_{Q,DOD}$ are lifetime degradation due to temperature, SOC, and DOD.

Each of the terms $L_{Q,T}$, $L_{Q,SOC}$ and $L_{Q,DOD}$, can be interpreted as a ratio $\left(\frac{\Delta L_x}{L_X}\right)$, where $\Delta L_x$ is the lifetime degradation due to a complete charging cycle during a day, and $L_X$ is the total battery lifetime if the charge cycle under evaluation were repeated until the battery's end of life (when energy capacity drops below 80 %).

Next, each of the terms above mentioned are defined. The term $L_{Q,T}$ relates the lifetime degradation with the temperature and charging time. $L_{Q,T}$ is proportional to charging power, this is because a high charging power corresponds to a high temperature. This is defined by

$$L_{Q,T} = \int_{t_{ch}} \frac{1}{nhy * l_y(T_{amb} + R_{th}|P_t(t)|)} dt + \frac{t_{max} - t_{ch}}{nhy * l(T_{amb})} \tag{27}$$

where $t_{ch}$, is the time of charging in hours, $t_{max}$ is the available time to perform a charging in hours, $nhy$ is the number of hours in a year. $T_{amb}$ is the ambient temperature setting in 25°C. $R_{th}$ is the thermal resistance of pack battery setting in $2°C/kW$. $P_t$ is the power of charging or discharging in (kW), and $l_y(T)$ is the total number of years the battery would last at that temperature, which is equal to $l_y(T) = ae^{\frac{b}{T}}$, where $a$ and $b$ are model parameters set as $\boldsymbol{a} = 3.73 \times 10^{-4}$ and $\boldsymbol{b} = 636$. [20]

The term $L_{Q,SOC}$ describes the battery lifetime degradation due to average SOC. It is defined by

$$L_{Q,SOC} = \frac{m_a * SOC_{avg} - d_a}{CF_{max} * y_p * nhy} \tag{28}$$

where $m_a = 1.6 \times 10^{-5}$ and $d_a = 6.4 \times 10^{-6}$ are model parameters defined in [20], $y_p = 15$ corresponds to the number of years projected of battery life, and $CF = 0.80$ is the capacity fading at the end of the battery life. We can notice that $L_{Q,SOC}$ is proportional to $SOC_{avg}$, therefore a high $SOC_{avg}$ reduces the battery life more than a moderated one.

The term $L_{Q,DOD}$ describes the battery lifetime degradation due to average *DOD*. It is calculated employing the concept of energy throughput [24], therefore $L_{Q,DOD}$, is the ratio between the energy throughput used in a complete charge cycle and the energy throughput used in all battery lifetime. It is defined by

$$L_{Q,DOD} = \frac{B \sum_{i=1}^{N_c} DOD_i}{N_l(DOD_{avg}) * DOD_{avg} * B} \tag{29}$$

where $DOD_i$ is the *i*th subcycle in a complete charge cycle, $N_c$ is the number of subcycles, $DOD_{avg}$ is *DOD* average, and $N_l(DOD)$ is the lifetime of battery in cycles for a given *DOD*. This is equal to $N_l(DOD) = \left(\frac{DOD}{145.71}\right)^{-\frac{1}{0.6844}}$, which is fitted according to the lithium-ion battery technology.

In brief, the model presented estimates the BEV battery degradation cost due to charging/discharging actions. Therefore, the battery degradation cost of public transport BEVs can be estimated. Also, the simplicity of the model allows easily to incorporate it into an optimization problem.

## 5. Differential Evolution

A *differential evolution* (DE) algorithm is presented to solve the problem formulated above. DE is a population-based and direct stochastic search algorithm, which was developed by Ken Price and Rainer Storn in 1996 [25, 26]. DE operates through similar computational steps as employed by a standard *evolutive algorithm (EA)*. However, DE employs difference of the population members to explore the objective function landscape, so it does not use any probability function to generate the offspring. In this way, DE uses a rather greedy and less stochastic approach in solving problems, which makes it more efficient to solve some kind of problems [27]. Table 4 shows the notation used in this section, and then the format of a solution and the DE algorithm steps are presented.

| Variable | Description |
|---|---|
| $\mathbf{X^U}$ | Vector containing the parameters for charge. |
| $\mathbf{X^a}$ | Vector containing the parameters for the route assignation. |
| $X_{i,G_e}$ | Vector containing parameters $i$ for generation $G_e$. |
| $V_{i,G_e}$ | Donor vector containing parameters $i$ for generation $G_e$. |
| $U_{i,G_e}$ | Trial vector containing parameters $i$ for generation $G_e$. |
| $N_p$ | Number of vector parameters. |
| $C_r$ | Crossover rate. |
| $m_v$ | Number of elements in a vector of parameters. |

Table 4: Notation for the differential evolution algorithm.

*5.1. Format of a Solution*

We must represent a possible solution in a format that DE can optimize. The possible solution of the problem of charge for EVs is represented by the valid profiles of assignment **a** and charge $\boldsymbol{u}^k$, for all $k \in K$. Both profiles **a** and $\boldsymbol{u}^k$, for all $k \in K$ are represented as follows

$$X^u = \left\{ u_x^1(1), \dots, u_x^1(N), \dots, u_x^{|K|}(1), \dots, u_x^{|K|}(N) \right\}$$

$$X^a = \left\{ a_1^1, \dots, a_1^{|K|}, \dots, a_{|H|}^1, \dots, a_{|H|}^{|K|} \right\}$$

where $X^u$ and $X^a$ are denoted as vectors of parameters of recharge and assignment, respectively. Thus, we define $X = X^u \cup X^a$ as a feasible solution of the problem of recharge for EVs, which is called the vector parameters $X$ in the DE context.

Similarly to most EA, the DE algorithm steps are: i) initialization of population; ii) mutation; iii) crossover; and iv) selection. Next, these are explained in detail.

*5.2. Initialization of Population*

A population is composed by a number of $N_p$ vector of parameters $X$, where each vector of parameters is also called an individual. We adopt the following notation for representing a vector of parameters $i \in N_P$ of the population at current generation $G_e$.

$$X_{i,G_e} = \left[x_{1,i,G_e}, x_{2,i,G_e}, \dots, x_{m_v,i,G_e}\right]$$

where $G_e = 0,1, \dots, G_{\max}$ denote the subsequent generations, $G_{\max}$ is the last generation, and $m_v$ is the dimension of a parameters vector.

The population of vector of parameters is generated by a random generator. Each vector of parameters is checked to meet the constraints, if it violates the constraints then it is rejected and it does not enter to population. $N_P$ is considered as a control variable of DE method, in this case, we use $N_P = 10m_v$ according to [27].

---
**Algorithm 1** Crossover subroutine
---
$L = 0$
**do**
$\quad L = L + 1$
**while** $(\text{rand}(0,1) \leq C_r$ and $L \leq m_v$
---

5.3. *Mutation*

The mutation allows to generate a donor vector $V_{i,G_e}$ by means of a differential operation. We define the differential operation as follows

$$V_{i,G_e} = X_{r_1,G_e} \; OR \; \left(X_{r_2,G_e} \; XOR \; X_{r_3,G_e}\right)$$

where $X_{r_1,G_e}, X_{r_2,G_e}$ and $X_{r_3,G_e}$ are three individual of population at generation $G_e$, the indices $r_1$, $r_2$, and $r_3$ are chosen mutually exclusive to diversify the mutation operation. These are randomly chosen from the range $[1, N_P]$.

The differential operation is based on adding the difference of two vectors to a vector of population [26]. In this case, the differential operation is defined by means of the exclusive disjunction (XOR). This is because, the parameters of vector X are binary variables and the XOR is a logical operation that outputs one if the binary inputs differ. In this way, the difference of vectors is drawn from the XOR operation, and the addition is result of the disjunction (OR) operation.

5.4. *Crossover*

The crossover operation allows the exchange of components between the donor vector $V_{i,G_e}$ and the target vector $X_{i,G_e}$. The result of crossover is the trial vector $U_{i,G_e} = [u_{1,i,G_e}, u_{2,i,G_e}, \dots, u_{m_v,i,G_e}]$ This operation is defined as follows

$$u_{j,i,G_e} = v_{j,i,G_e} \; for \; n_v + 1 \leq j \leq n_v + L - 1$$
$$u_{j,i,G_e} = x_{j,i,G_e} \; otherwise$$

where $n_v$ is randomly chosen among $[1, m_v]$, and $L$ is drawn from $[1, m_v]$ according to Algorithm 1 where $C_r \in [0,1]$ is called the crossover rate and it constitutes a control variable of the DE method.

$C_r$ must be considerably lower than one (e.g., 0.3). But, if no convergence can be achieved, $C_r$ can be chosen in the range [0.5, 1] according to [27].

*5.5. Selection*

The selection operation determines whether the target vector $X_{i,G_e}$ or the trial vector $U_{i,G_e}$ survives to the next generation $G_e + 1$. The selection operation is defined as follows

$$X_{i,G_e+1} = U_{i,G_e} \quad if\ f(U_{i,G_e}) \leq f(X_{i,G_e})$$
$$X_{i,G_e+1} = X_{i,G_e} \quad otherwise$$

where $f(.)$ is the objective function (25) to be minimized.

The selection operation means that if the trial vector $U_{i,G_e}$ yields an equal or upper value of the objective function, then it replaces the target vector in the next generation, otherwise the target vector $X_{i,G_e}$ is retained in the population at generation $G_e + 1$.

In this manner, the steps *mutation* to *selection* are cycled until the maximum generation $G_{\max}$, is reached, which is outlined in Algorithm 2.

---

**Algorithm 2** DE algorithm

    **Inputs:** Read values of the control parameters: $C_r$, $N_P$
    **Initialization of population ($P_{G_e}$):** $G_e = 0$, $P_G = \{X_{1,G_e}; \ldots; X_{N_P,G_e}\}$
    **for** $G_e = 1$ to $G_{max}$ **do**
        **for** $i = 1$ to $N_P$ **do**
            **Mutation:** Generate a donor vector $V_{i,G}$
            $V_{i,G_e} = X_{r2,G_e}\ \text{OR}\ (X_{r2,G_e}\ \text{XOR}\ X_{r3,G_e})$
            **Crossover:** Generate a trial vector $U_{i,G_e}$
            $u_{j,i,G_e} = v_{j,i,G_e}$ for $n_v + 1 \leq j \leq n_v + L - 1$
            $u_{j,i,G_e} = x_{j,i,G_e}$ otherwise
            **Selection:** Evaluate the trial vector $U_{i,G_e}$
            $X_{i,G_e+1} = U_{i,G_e}$ if $f(U_{i,G_e}) \leq f(X_{i,G_e})$
            $X_{i,G_e+1} = X_{i,G_e}$ otherwise
        **end for**
    **end for**

---

## 6. Case Study

An airport-shuttle service (SA) with electric vehicles is presented as case study. It consists on determining the route assignment and the scheduling of recharge for an EV fleet that will provide the SA service near El Dorado Airport in Bogotá, where the SA will take place. Four aspects concerning the case of study are described: i) the operation zone; ii) BEV characteristics and data logging; iii) data processing; and iv) the transport demand.

*6.1. Operations zone*

The operation zone proposed for the SA is comprised within two districts of Bogotá near El Dorado airport, which is also the north-westernmost node; a Shopping Mall is the southernmost node; and a public recharge station (RS) as the easternmost node.

Within this area, a total of 54 nodes were used as a reference to determine the routes; 6 of which act as the main nodes used to build the graph (see Fig. 4 and Table 5): 1) hotel, 2) private RS with $r_2 = 3kW$; $\eta_2 = 0:9$, 3) public RS with $r_3 = 6kW$; $\eta_3 = 0:9$, 4) shopping mall, 5) airport whereabout 1, 6) airport whereabout 2, and 7) the bus terminal. It is assumed that the OC and the private RS are located at the hotel.

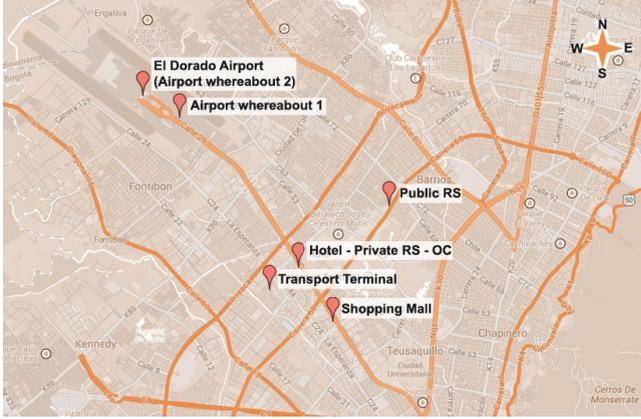

Figure 4. Main nodes of the operations area.

|  | **Airport 1** | **Airport 2** | **Hotel** | **Public RS** | **Mall** | **Terminal** |
|---|---|---|---|---|---|---|
| **Airport 1** | 0 | 0.94 | 5.92 | 9.55 | 7.62 | 6.69 |
| **Airport 2** | 2.94 | 0 | 5.86 | 9.54 | 7.6 | 6.66 |
| **Hotel** | 6.2 | 7.11 | 0 | 5.5 | 1.79 | 3.33 |
| **Public RS** | 6.99 | 7.87 | 4.21 | 0 | 4.21 | 5.9 |
| **Mall** | 7.74 | 8.62 | 2.61 | 4.9 | 0 | 4.59 |
| **Terminal** | 7.48 | 8.37 | 2.1 | 6.03 | 3.25 | 0 |

Table 5: Distance matrix [km].

### 6.2. Test vehicle characteristics

The characteristics of the vehicle used during the measurements are summed up in Table 6. Due to the necessity to perform the study prior to the arrive of the BEVs, a test vehicle was selected so that its characteristics where as close as possible to the ones of an BEVs. Having this consideration in mind, it was determined that the vehicle had to be a saloon with similar dimensions (therefore similar $C_D$ and frontal area) to that of an BEV destined for use as a taxi. It was also taken into account that the power to weight ratio of the vehicle had to be near to that of such BEV. Finally, since BEVs tend to have slightly lower power figures and more mass than an internal combustion engine vehicle of the same size, the test vehicle had to be fitted with extra weight to fulfill the requirements of the power to weight ratio.

| **Parameter** | **Value** |
|---|---|
| m[kg] | 1 312 |
| A[m$^2$] | 1.86 |
| $C_D$ | 0.32 |
| $f_r$ | 0.0117 |
| $\hat{\eta}$ | 0.9 |
| Overall length [m] | 4.32 |
| Overall width [m] | 1.69 |

Table 6: BEV characteristics.

*6.3. Data logging and data processing*

The characteristics of the GPS unit used to collect the data are summed up in Table 7. Two main issues arise within the speed profile data obtained from the GPS device that should require a stage of signal conditioning (see Fig. 5): noise due to poor quality signal and outlier points due to a complete loss of communication between the GPS unit and the satellites.

| Parameter | Value |
|---|---|
| Log Rate [Hz] | 100 |
| Speed Resolution $\left[\frac{km}{h}\right]$ | 0.01 |
| Speed Accuracy $\left[\frac{km}{h}\right]$ | 0.1 |
| Distance $Resolution$ [$m$] | 0.01 |
| Distance Accuracy [%] | 0.05 |

Table 7: GPS unit characteristics.

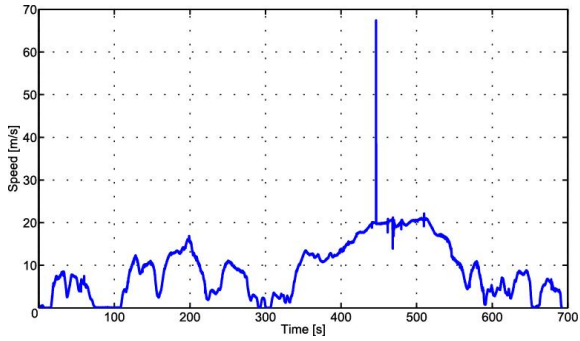

Figure 5. Raw speed profile.

In order to obtain a smooth speed profile two methods were considered: Kalman filtering described in [28] and Savitzky-Golay filtering, described in [29, 30]. In Fig. 6), a comparison between each of the two filters against the original data is presented. It can be seen that, for this case of study, the Savitzky-Golay filter (see Fig. 6(left)) is more effective than the Kalman filter (see Fig. 6(right)).

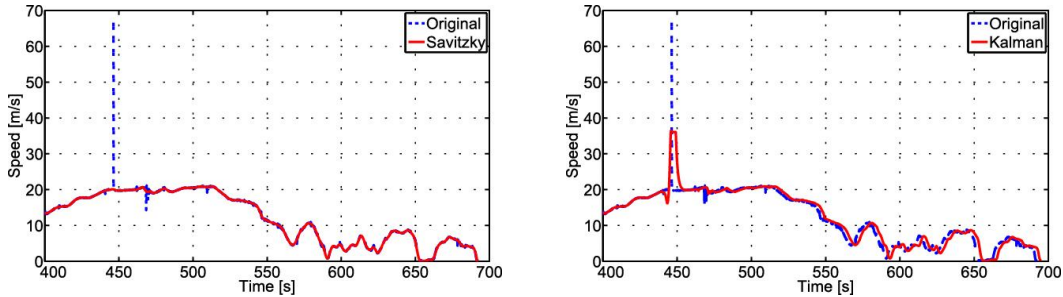

Figure 6: Savitzky-Golay filter (left) and Kalman filter (right) comparison.

*6.4. Transport demand*

In addition, it is defined an illustrative transport demand, which consists of one passenger departing from the airport whereabout 1 and two passengers departing from the airport whereabout 2 each hour, from 8:00 to 16:00. The passengers must pay a fee equal to $5 and they must be picked up within a time window of 30 minutes. Moreover, it is assumed that the passengers must be delivered in the shopping center, the terminal or the hotel according to their needs.

*6.5. Results*

Finally, results on the energy consumed by the vehicle during a trip along each path are calculated. This is achieved by introducing each filtered velocity profile (see Fig. 6) separately into the longitudinal dynamics model, described in section 2.1. The results are shown in the energy matrix (see Table 8), in which each row represents a departing node, and each column represents an arrival node.

|  | **Airport 1** | **Airport 2** | **Hotel** | **Public RS** | **Mall** | **Terminal** |
|---|---|---|---|---|---|---|
| **Airport 1** | 0 | 0.128 | 0.574 | 1.073 | 0.79 | 0.719 |
| **Airport 2** | 0.285 | 0 | 0.567 | 1.076 | 0.634 | 0.656 |
| **Hotel** | 0.593 | 0.773 | 0 | 0.658 | 0.183 | 0.429 |
| **Public RS** | 0.688 | 0.828 | 0.51 | 0 | 0.443 | 0.727 |
| **Mall** | 0.805 | 0.876 | 0.321 | 0.497 | 0 | 0.536 |
| **Terminal** | 0.839 | 0.925 | 0.28 | 0.72 | 0.407 | 0 |

Table 8: Energy consumption matrix [kWh].

**7. Simulation Results**

First of all, the proposed EVRP has been solved using the optimization tool XPRESS. In this way, it was found that it takes 9 routes to satisfy the transport demand, where each route is optimal regarding to the energy consumption. The optimization tool results show that each route has an energy consumption of 1.66kWh approx. and all routes take place during the time interval [7:30 - 15:30], one every hour.

Then, the charging and routes assignment has been planned during a programming horizon from 7:00 to 7:00 split in half hour intervals. To illustrate possible circumstances, four scenarios (SC) under different conditions were considered. In SC 1, the charging is only allowed in the RS 2 with an energy tariff of two level structure, and the battery degradation cost is not considered. In SC 2, the conditions of SC 1 are maintained, with the exception that the battery degradation cost is now considered. SC 3 and 4, are based on SC 1 and 2 respectively, but the charging in the RS 3 with different energy tariffs is allowed. Next, the results for each proposed scenario are shown.

Figs. 7(a) and 7(b) show the charging and routes assignment obtained in SC 1 and 2 using the DE method. Fig. 7(a) shows that the routes are assigned arbitrarily between the BEVs and the recharging actions (RAs) are scheduled later, which is encouraged by the energy low cost at that hours. On the other hand, in Fig. 7(b), it was found that the routes are assigned almost uniformly between the BEVs. Also, the BEVs SOC describes a pattern travel-charge, i.e., the BEVs travel along few routes and after charge a little of energy. It has been noticed that the charging pattern and the routes assignment shown in 7(b) allows to reduce the degradation cost because the DOD is kept at low values, which is the dominant component of battery degradation cost. Consequently, it was found that while the estimated battery life for BEV 1 in SC 1 is approx. 6400 cycles, the other one in Fig. 7(b) is approx. 8400 cycles, a difference of 2000 cycles which is comparable to 5 years of operation as a shuttle service vehicle.

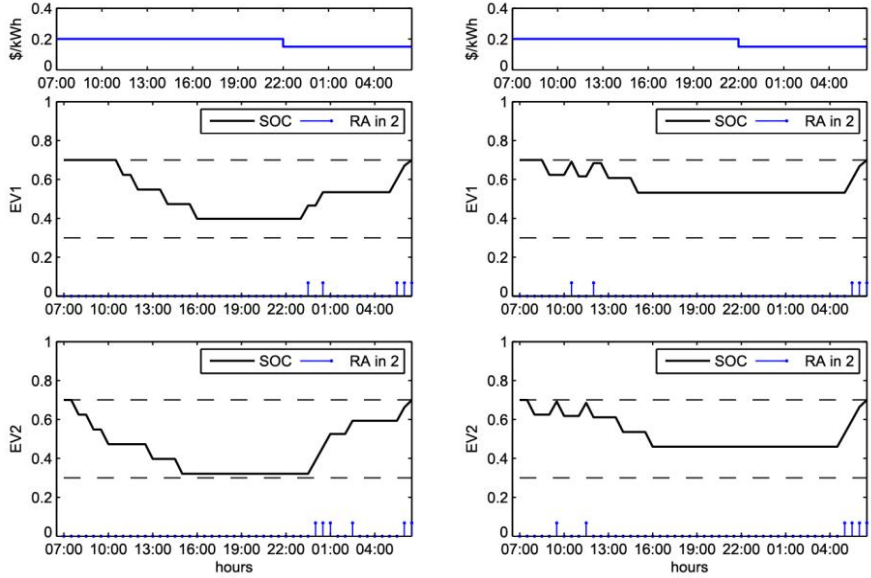

Figure 7: Scheduling of charge for BEV 1 and BEV 2 operating in scenarios 1 and 2.

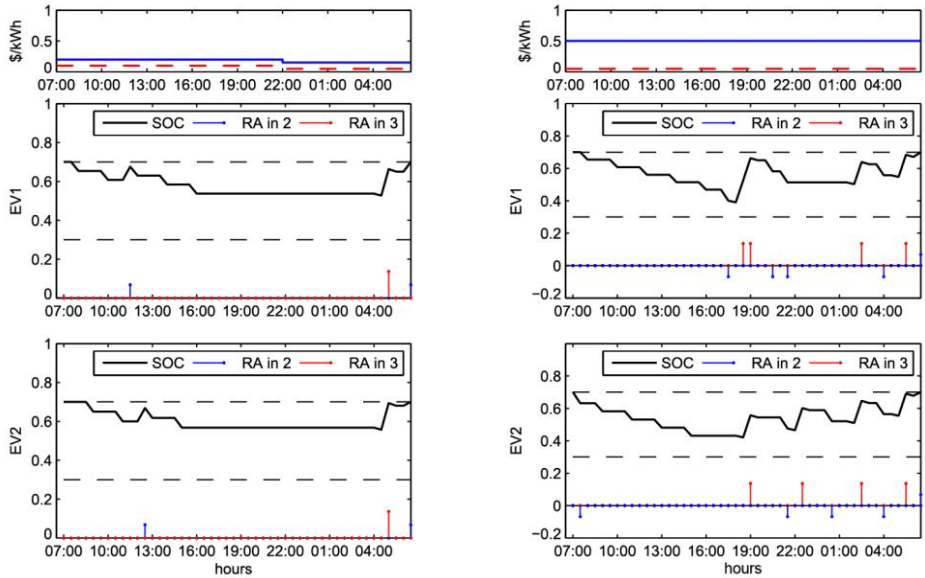

Figure 8: Scheduling of charge for BEV1 and BEV2 operating in scenarios 3 and 4.

Figs. 8(a) and 8(b) show the scheduling of charge obtained in SC 3 and 4. Fig. 8(a) shows that the V2B operation in the hotel is not scheduled. By contrast, Fig. 8(b) shows that several RAs in RS 3 are scheduled and the V2B operation is performed. It is noticed that the energy tariff in scenario 3 is not profitable because the profit of allowing a V2B operation is overcame by the battery degradation cost. However, when the difference between energy tariff in RS 2 and 3 gets higher (i.e., $0.5) then the V2B operation is profitable. In this manner, the operation cost in SC 3 is $4.83, while the cost in SC 4 is $1.04.

## 8. Conclusions

A methodology to plan the charging and the routes assignment for a BEVs fleet has been presented. This methodology considers the search of optimal routes and the minimization of operation costs. It has been found that the scheduling of charge and routes assignment have effects on the battery lifetime. A pattern for charging has been obtained, which allows increasing the battery lifetime. In addition, it is noticed that this pattern is slightly different to the one obtained when only the charging cost is considered.

Furthermore, some conditions to perform a V2B operation have been presented. These are related with the battery technology, the battery degradation model, and the energy tariff. It has been found that the battery with lithium-ion technology studied in this paper does not meet the requirements to provide the V2B operation since its degradation cost is large. Hence, the V2B operation is only profitable for BEVs owner when the difference of the energy tariff is approx. $0.5. However, this energy tariff does not correspond to a real scenario.